\documentclass[11pt]{article}
\usepackage[a4paper]{geometry}
\usepackage[english]{babel}

\usepackage{amsmath, accents}
\usepackage{amsfonts}
\usepackage{amstext}
\usepackage{amssymb}
\usepackage{amsthm}
\usepackage{amscd}

\usepackage[pagebackref,draft=false]{hyperref}
\hypersetup{colorlinks,
linkcolor=myrefcolor,
citecolor=mycitecolor,
urlcolor=myurlcolor}

\usepackage[capitalize]{cleveref}
\usepackage{cite}
\usepackage{caption}
\usepackage{etaremune}

\usepackage{xcolor}
\definecolor{myurlcolor}{rgb}{0,0,0.4}
\definecolor{mycitecolor}{rgb}{0,0.5,0}
\definecolor{myrefcolor}{rgb}{0.5,0,0}
\usepackage{graphicx}
\usepackage{tikz}
\usepackage{tikz-cd}
 \usetikzlibrary{decorations.pathmorphing}

\usepackage{mathrsfs}
\newcommand{\be}{\begin{equation}}
\newcommand{\ee}{\end{equation}}
\newcommand{\bea}{\begin{eqnarray}}
\newcommand{\eea}{\end{eqnarray}}
\newcommand{\vsp}{\vspace{0.4cm}}
\newcommand{\grit}[1]{{\bfseries {\itshape {#1}}}}

\newcommand{\ra}{\rightarrow}

\newcommand{\stav}{\mathscr{V}}

\newcommand{\dd}{{\rm d}}

\title{Covariant Variational Evolution and Jacobi Brackets: Fields}
\date{}

\author{F. M. Ciaglia$^{1,7}$ \href{https://orcid.org/0000-0002-8987-1181}{\includegraphics[scale=0.7]{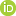}}, F. Di Cosmo$^{2,3,8}$ \href{https://orcid.org/0000-0003-0256-5913}{\includegraphics[scale=0.7]{ORCID.png}}, A. Ibort$^{2,3,9}$ \href{https://orcid.org/0000-0002-0580-5858}{\includegraphics[scale=0.7]{ORCID.png}}, \\ G. Marmo$^{4,5,10}$ \href{https://orcid.org/0000-0003-2662-2193}{\includegraphics[scale=0.7]{ORCID.png}}, L. Schiavone$^{3,4,6,11}$  \href{https://orcid.org/0000-0002-1817-5752}{\includegraphics[scale=0.7]{ORCID.png}} \\
\footnotesize{$^{1}$\textit{ Max Planck Institute for Mathematics in the Sciences, Leipzig, Germany}} \\
\footnotesize{$^{2}$\textit{ ICMAT, Instituto de Ciencias Matem\'{a}ticas (CSIC-UAM-UC3M-UCM)}} \\
\footnotesize{$^{3}$\textit{Depto. de Matem\'aticas, Univ. Carlos III de Madrid, Legan\'es, Madrid, Spain}} \\
\footnotesize{$^{4}$\textit{ INFN-Sezione di Napoli, Naples, Italy}} \\
\footnotesize{$^{5}$\textit{ Dipartimento di Fisica ``E. Pancini'', Universit\`a di Napoli Federico II,  Naples, Italy}} \\
\footnotesize{$^{6}$\textit{ Dipartimento di Matematica e Applicazioni "Renato Caccioppoli", Università di Napoli Federico II, Napoli, Italy}} \\
\footnotesize{$^{7}$\textit{ e-mail: \texttt{florio.m.ciaglia[at]gmail.com} and \texttt{ciaglia[at]mis.mpg.de}}} \\
\footnotesize{$^{8}$\textit{ e-mail: \texttt{fcosmo[at]math.uc3m.es}}} \\
\footnotesize{$^{9}$\textit{ e-mail: \texttt{albertoi[at]math.uc3m.es}}} \\
\footnotesize{$^{10}$\textit{ e-mail: \texttt{marmo[at]na.infn.it}}} \\ 
\footnotesize{$^{11}$\textit{ e-mail: \texttt{luca.schiavone[at]unina.it}}}  
}

\begin{document}

\maketitle

\begin{abstract}
The analysis of the covariant brackets on the space of functions on the solutions to a variational problem in the framework of contact geometry initiated in the companion letter \cite{C-DC-I-M-S-2020-01} is extended to the case of the multisymplectic formulation of the free Klein-Gordon theory and of the free Schr\"{o}dinger equation.

\end{abstract}

\section{Introduction}

In the companion letter \cite{C-DC-I-M-S-2020-01}, the problem of the definition of a covariant bracket on the space of functions on the solutions of a suitable variational principle has been addressed from the point of view of the geometry of contact manifolds.
Specifically, in the case of non-relativistic Hamiltonian mechanics, and of the relativistic particle, it was proved how the covariant bracket may be read in terms of a Poisson subalgebra of the algebra of smooth functions on a contact manifold endowed with its natural Jacobi bracket\cite{AsoCiagliaDCosmoIbortMarmo2017-Covariant_Jacobi_brackets,AsoCiagliaDCosmoIbort2017-Covariant_brackets,C-C-M-2018}.
In this letter, we push further this analysis by considering the case of Klein-Gordon theory, and the case of the multisymplectic formulation of Schr\"{o}dinger equation.
The main difference with the cases considered in  the companion letter\cite{C-DC-I-M-S-2020-01} is given by the fact that these systems describe partial differential equations.
Consequently, we will see that it is possible to describe them in terms of two different but equivalent action principles taking place on two different spaces of sections.

On the one hand, it is possible to formulate an action principle for sections of a suitable finite-dimensional fiber bundle on the spacetime manifold in the spirit of the multisymplectic formalism\cite{CarinCrampIbort1991-Multisymplectic,IbSpiv2017-Covariant_Hamiltonian_boundary, KijowskiTulczyjew1979, BinzSniatyckiFisher1988, GoldschmidtSternberg}.
In this case, it is not possible to express the covariant bracket  in terms of a Poisson subalgebra of a suitable contact manifold (see, for instance,  \cite{Cr87,Cr88,MarsdMongMorrThom1986-Covariant_Poisson_bracket} for historical attempts in writing a covariant bracket for Field Theory within the canonical formalism and \cite{ForgRomero2005-Covariant_poisson_brackets,Fredenhagen2015-Algebraic_quantum_field_theory}, and references therein, for a more modern geometric approach).
However, this point of view may be considered as the starting point for  the generalization of the notion of contact manifold and contact structure to framework which is more suitable for field theories along the lines presented in \cite{Vinogradov-Contact_Monge}. 
In this sense, we expect that an approach in terms of multicontact geometry could help for a better understanding of the covariant formalism, in the same way as multisymplectic formalism has generalized symplectic geometry, and leave  the discussion of these aspects to a future work.

On the other hand, it is possible to formulate an action principle for sections of a suitable \grit{infinite-dimensional} fiber bundle over the time manifold $\mathbb{R}$, in analogy with the particle case.
In this context, the infinite-dimensional nature of the fiber bundle will make possible to read the Poisson bracket on the space of functionals on the solutions of the action principle in terms of a Poisson subalgebra of the algebra of smooth functions endowed with a Jacobi bracket.

\section{Multisymplectic formulation of free Klein-Gordon theory}\label{sec: KG}

We consider the covariant Hamiltonian description of Klein-Gordon theory on the Minkowski spacetime $\mathcal{M} = (\mathbb{R}^4,\eta)$, where $\eta$ is the metric tensor 
\be
\eta = \eta_{\mu \nu} \dd x^{\mu} \otimes \dd x^{\nu} = -\dd x^0 \otimes \dd x^0 + \delta_{jk}\dd x^j \otimes \dd x^k ,
\ee
and $(x^{\mu})$ is a global set of Cartesian coordinates.
Let $\pi_0\colon E=\mathbb{R}\times\mathcal{M}\,\rightarrow \,\mathcal{M}$ be a bundle with $\pi_{0}$ the standard projection on the second factor.
This bundle represents the analogue of the extended configuration space  $\mathcal{Q}\times \mathbb{R}$ of non-relativistic Hamiltonian mechanics considered in the companion letter \cite{C-DC-I-M-S-2020-01}.
Klein-Gordon fields are sections $\phi$ of the bundle $\pi_0\colon E=\mathbb{R}\times\mathcal{M}\,\rightarrow \,\mathcal{M}$.

The extended phase space $\mathbf{T}^{*}\mathcal{Q}\times \mathbb{R}$ of non-relativistic particle mechanics is here replaced by the so called covariant phase space $\mathcal{P}$ which is a fibre bundle over both $E$ and the spacetime $\mathcal{M}$ \cite{CarinCrampIbort1991-Multisymplectic,IbSpiv2017-Covariant_Hamiltonian_boundary, BinzSniatyckiFisher1988, KijowskiTulczyjew1979}.
Technically speaking, $\mathcal{P}$ is the (affine) dual bundle of the first jet bundle $J^1(E)$ that can be identified with the space of 1-semibasic $4$-forms on $E$.
In the Klein-Gordon case on Minkowski spacetime we are considering, $\mathcal{P}$ is diffeomorphic to $E\times\mathbb{R}^{4}$. 
If $(x^{\mu})$, $(u,x^{\mu})$, and $\left( \rho^{\mu},u,x^{\mu}  \right)$   are, respectively, Cartesian coordinates on $\mathcal{M}$, bundle coordinates on $E$, and bundle coordinates on $\mathcal{P}$, the projection $\pi_{E}\colon \mathcal{P}\,\rightarrow \, E$ is locally given by
\begin{equation}
 \pi_{E}(\rho^{\mu},u,x^{\mu})  = (u,x^{\mu}) \,,
\end{equation}  
while the projection $\pi\colon \mathcal{P}\,\rightarrow \, \mathcal{M}$ is    given by
\begin{equation}
\pi(\rho^{\mu},u,x^{\mu}) = (x^{\mu})\,. 
\end{equation}  

Let $\mathscr{F}_{\mathcal{P}}$ denote the space of sections $\chi\colon \mathcal{M}\,\rightarrow\,\mathcal{P}$  of the form:
\begin{equation}
\chi(x^{\mu}) = (x^{\mu}, \phi(x^{\mu}), P^{\nu}(x^{\mu}))\,, \quad \mu,\nu = 0,1,2,3\,.
\end{equation}
Elements in $\mathscr{F}_{\mathcal{P}}$ will be the fields of the Hamiltonian theory where $P^\nu$ are identified with the momenta fields of the theory.
A rigorous analytical framework can be provided by introducing some regularity conditions on the various spaces of fields, for instance, we may consider the fields $\phi$ in the Sobolev space $\mathcal{V}= \mathcal{H}^1(\mathcal{M},\mathrm{vol}_{\mathcal{M}})$ of square integrable functions on $\mathcal{M}$ with respect to the Lebesgue measure on $\mathcal{M}$, and the momenta fields $P = (P^{\mu})$ in the Hilbert space $\mathcal{W} = L^2(\phi^*\mathcal{P})$ of square integrable sections of the pullback of the bundle $\mathcal{P} \to E$ along $\phi$.
With these choices, the space $\mathscr{F}_{\mathcal{P}}$  becomes the Hilbert space $\mathscr{F}_{\mathcal{P}}\,=\,\mathcal{V}\,\oplus\,\mathcal{W}$.

A variation for $\chi\in\mathscr{F}_{\mathcal{P}}$, commonly denoted as $\delta \chi$, is a tangent vector at  $\chi$, which may be identified with a vector field $U_{\chi}$ along $\chi$ on $\mathcal{P}$, which is vertical with respect to the fibration $\pi\colon\mathcal{P}\ra\mathcal{M}$.
The tangent space $\mathbf{T}_{\chi}\mathscr{F}_{\mathcal{P}}$ is given by all variations $\delta\chi_U = U_{\chi}$.
In the following, it will be useful to extend $U_{\chi}$ to a vertical vector field $\widetilde{U}$ on a neighbourhood of the image of $\chi$ within $\mathcal{P}$, as:
\begin{equation}
\tilde{U} = U_{\phi} \frac{\partial }{\partial u} + U_{P}^{\mu}\frac{\partial }{\partial \rho^{\mu}}\,.
\end{equation}


Now, we pass to describe the dynamics in terms of the (Schwinger-Weiss) action principle.
Given a Hamiltonian function $H\colon \mathcal{P}\,\rightarrow\,\mathbb{R}$ and volume form $\mathrm{vol}_\mathcal{M}  = \dd x^0\wedge \dd x^1 \wedge \dd x^2 \wedge \dd x^3$ on the base manifold $\mathcal{M}$, we consider the  4-form\footnote{It is possible to define this form in an intrinsic way, see for instance \cite{IbSpiv2017-Covariant_Hamiltonian_boundary}.} $\theta_H$  on $\mathcal{P}$ given by
\begin{equation}
\theta_H = \rho^{\mu}\dd u \wedge i_{\frac{\partial}{\partial x^{\mu}}}\mathrm{vol}_\mathcal{M}  - H \mathrm{vol}_\mathcal{M}\,.
\end{equation}   
In the particular instance of  free dynamics, the Hamiltonian function is
\be\label{eqn: relativistic Hamiltonian}
H = \frac{1}{2} \left( \eta_{\mu \nu} \rho^{\mu}\rho^{\nu} - \mathfrak{m}^2 u^2 \right)\,. 
\ee

The action functional $S\colon\mathscr{F}_{\mathcal{P}}\, \rightarrow\, \mathbb{R}$, of the theory can be written as:
\begin{equation}
S[\chi] = \int_{\mathcal{M}}\chi^*\left( \theta_H \right) = \int_{\mathcal{M}}\left( P^{\mu}\partial_\mu \phi - H \right) \mathrm{vol}_\mathcal{M}\,.
\label{action_klein-gordon}
\end{equation}
Given  $U_{\chi}\in\mathbf{T}_{\chi}\mathscr{F}_{\mathcal{P}}$, the variation  $\mathrm{d}S[\chi](U_{\chi})$ of $S$ is 
\be
\begin{split}
\mathrm{d}S[\chi](U_{\chi})&=  \int_{\mathcal{M} }\chi^*\left( \mathrm{L}_{\tilde{U}}\theta_H \right)  = \int_{\mathcal{M} }\chi^*\left( i_{\tilde{U}}\dd \theta_H \right) + \int_{\partial \mathcal{M} } \chi_{\partial\mathcal{M}}^*\left( i_{\tilde{U}}\theta_H \right),
\end{split}
\ee
where $\tilde{U}$ is any extension of $U_{\chi}$,
and, if $\partial\mathcal{M}=\emptyset$, we get:
$$
\mathrm{d}S[\chi](U_{\chi}) = \,\int_{\mathcal{M} }\chi^*\left( i_{\tilde{U}}\dd \theta_H \right) = \int_{\mathcal{M} } \left(\frac{\partial \phi}{\partial x^\mu} - \frac{\partial H}{\partial P^\mu} \right) U_P^\mu +  \left(\frac{\partial P^\mu}{\partial x^\mu} + \frac{\partial H}{\partial \phi} \right) U_\phi^\mu \,  \mathrm{vol}_\mathcal{M}\, .
$$
The Schwinger-Weiss action principle states that the variations of the action depend solely on the variations of the fields at the boundary, hence the actual dynamical configurations of the fields of the theory must satisfy the Euler-Lagrange equations:
$$
\frac{\partial \phi}{\partial x^\mu} = \frac{\partial H}{\partial P^\mu} \, ,  \qquad  \frac{\partial P^\mu}{\partial x^\mu} = - \frac{\partial H}{\partial \phi} \, ,
$$
which can be geometrically interpreted as the zeroes of a 1-form $\mathbb{EL}$ on the space of fields $\mathscr{F}_{\mathcal{P}}$ given by:
\begin{equation}\label{eqn: Schwinger-Weiss action principle 1}
\mathbb{EL}_{\chi}(U_{\chi}) := \int_{\mathcal{M} }\chi^*\left( i_{\tilde{U}}\dd \theta_H \right) = 0 \,,\quad \forall U_{\chi}\in\mathbf{T}_{\chi}\mathscr{F}_{\mathcal{P}}\, .
\end{equation}
We note that such a geometrical reformulation of the Schwinger-Weiss variational principle should be used to introduce a Quantum Action Principle in the groupoid reformulation of Schwinger's algebra of selective measurements which has been recently proposed by some of the authors (see \cite{C-I-M-2018,C-I-M-02-2019,C-I-M-03-2019,C-I-M-05-2019,C-DC-I-M-2020,C-DC-I-M-02-2020} for more details).

We will denote by $\mathcal{EL}_{\mathcal{M}}\subset \mathscr{F}_{\mathcal{P}}$ the space of solutions of Euler-Lagrange equations, that is:  
\begin{equation}
\mathcal{EL}_{\mathcal{M}}\,:=\,\left\{\chi\in \mathscr{F}_{\mathcal{P}}\,\colon\,\mathbb{EL}_{\chi}(U_\chi) = 0\,,\quad \forall U_{\chi} \in \mathbf{T}_{\chi}\mathscr{F}_{\mathcal{P}}\right\}.
\end{equation} 
Hence, we obtain the Euler-Lagrange equations  (also known as the de Donder-Weyl equations)  for the Klein-Gordon theory:
\be
\frac{\partial \phi}{\partial x^{\mu}} = \eta_{\mu \nu} P^{\nu}\,,\qquad\, \frac{\partial P^{\mu}}{\partial x^{\mu}} =  \mathfrak{m}^2 \phi\,.
\ee
Analogously to what happens in non-relativistic Hamiltonian  mechanics   \cite{C-DC-I-M-S-2020-01}, if we select a codimension-one, spacelike submanifold $\Sigma \subset \mathcal{M}$, for instance  
\be
\Sigma=\{m\in\mathcal{M}\,|\,x^{0}(m)= \tau^0 \},
\ee
and denote  by $i_{\Sigma}:\Sigma\ra\mathcal{M}$ the canonical immersion of $\Sigma$ in $\mathcal{M}$, the space $\mathcal{EL}_{\mathcal{M}}$ is equipped with a 2-form $\Omega_{\chi}$ given by:
\begin{equation}\label{eqn: omega on solutions of de Donder Weyl for KG}
\Omega_{\chi} (U_{\chi},V_{\chi}) = \int_{\Sigma}i_{\Sigma}^*\left( \chi^*\left( i_{\tilde{V}}i_{\tilde{U}} \dd \theta_H \right) \right) = \int_{\Sigma} i_{\Sigma}^*\left( \delta P^{0}_U\,  \delta\phi_V  - \delta\phi_U \, \delta P^{0}_V  \right) \mathrm{vol}_\Sigma \ , \nonumber
\end{equation}
with $U_\chi = \delta \phi_U \partial /\partial u + \delta P^\mu_U \partial /\partial \rho^\mu$ and $V_\chi$ similarly.
Following the procedure outlined in \cite{C-DC-I-M-S-2020-01}, provided that $\chi \in \mathcal{EL}_{\mathcal{M}}$, it is possible to prove that $\Omega_{\chi}$  is independent of the choice of $\Sigma$, and it defines a canonical symplectic structure on the space of solutions of Klein-Gordon equation.

\vsp

Now, we will provide an evolution description of Klein-Gordon theory in terms of a vector field on an infinite-dimensional manifold.
The main idea is to provide a framework in which the role of a contact structure is manifestly evident, and which allows us to write the Poisson bracket associated with $\Omega$ in terms of the Jacobi bracket defined by such structure as done in the companion letter\cite{C-DC-I-M-S-2020-01}  for the case of non-relativistic Hamiltonian mechanics and the relativistic particle.

To define the infinite-dimensional manifold which will be the carrier space of the dynamics, we fix a space-like, codimension-one submanifold $\Sigma\subset\mathcal{M}$ as before.
Then, we consider the pullback bundle $i_{\Sigma}^{*}\mathcal{P}$ whose sections $\sigma$ are just compositions of sections $\chi$ of $\mathcal{P}$ with $i_{\Sigma}$, $\sigma = \chi \circ i_\Sigma$ and, in local coordinates, we have:
$$
\sigma(x^{j})=\left(x^{j},\tau^0, \varphi(x^{j}), \, p(x^{j}) ,\,  \beta^k(x^{j}) \, \right),
$$
where
\be
\varphi(x^{j}) =(\phi(x^{j},\tau^{0})) |_{\Sigma}\, , \quad 
p(x^{j}) =(P^{0}(x^{j},\tau^{0})) |_{\Sigma} \, , \quad 
\beta^k(x^{j}) =(P^k(x^{j},\tau^{0})) |_{\Sigma}\,.
\ee 
As before, we will focus on those sections of $i_{\Sigma}^{*}\mathcal{P}$ satisfying some additional regularity conditions.
Specifically, we will consider $\varphi\in\stav_\Sigma$, the space of sections of Sobolev class 1 of the bundle $i_\Sigma^*E$,  $p\in\mathscr{W}_\Sigma =\mathcal{L}^2(\Sigma, \mathrm{vol}_{\Sigma})$, and $\beta = (\beta^{j}) \in\mathscr{B}_\Sigma$, the space of square integrable sections of the tangent bundle $\mathbf{T}\Sigma$, so that we have the Hilbert space $ \mathcal{F}_{\Sigma}$ of fields at $\Sigma$ given by $\mathcal{F}_{\Sigma}\,:=\,\stav_\Sigma \,\oplus\,\mathscr{W}_\Sigma\,\oplus\,\mathscr{B}_\Sigma$.

Then, we consider the Hilbert bundle $\tau\colon\mathcal{F}_{\Sigma}\times\mathbb{R}\ra\mathbb{R}$, where $\tau$ is the projection on the second factor, which plays the role of the extended phase space in \cite{C-DC-I-M-S-2020-01}, and we denote by $\Gamma(\mathcal{F}_\Sigma)$ the space of sections of this bundle.
The key observation is that, under suitable regularity conditions, the sections in $\Gamma(\mathcal{F}_\Sigma)$, i.e., curves $\sigma \colon \mathbb{R} \to \mathcal{F}_{\Sigma}$, are in one-to-one correspondence with fields in $\mathscr{F}_{\mathcal{P}}$.
Indeed, given $\chi\in\mathscr{F}_{\mathcal{P}}$, we can define the section $\sigma_{\chi}\in\Gamma$ setting
\be
\sigma_{\chi}(s)\,:=\,\left(s,x^{j}, \tau^{0},\varphi_{s}(x^{j}),p_{s}(x^{j}),\beta^{k}_{s}(x^{j})\right),
\ee
where $s=x^{0}$, and 
\be
\varphi_{s}(x^{j}) =\phi(x^{j},s) \, , \qquad 
p_{s}(x^{j}) =P^{0}(x^{j},s)  \, , \qquad 
\beta^{k}_{s}(x^{j}) = P^k(x^{j},s) \,.
\ee 
On the other hand, given $\sigma \in\Gamma(\mathcal{F}_\Sigma)$, we can define $\chi_{\sigma}\in\mathscr{F}_{\mathcal{P}}$ by reading the previous equations in the other direction.
The regularity conditions for $\sigma\in\Gamma(\mathcal{F}_\Sigma)$ alluded to above are precisely those assuring that $\phi(x^{j},s):=\varphi_{s}(x^{j})$ is in $\mathcal{V}$, and that $(P^{0}(x^{j},s):=p_{s}(x^{j}),P^k(x^{j},s):=\beta^{k}_{s}(x^{j}))$ are in $\mathcal{W}$.

\vsp
A variation $U_{\gamma}$ at $\gamma\in \Gamma (\mathcal{F}_\Sigma)$ is then a  vector field along $\gamma$  which is vertical  with respect to the fibration $\tau\colon\mathcal{F}_{\Sigma}\times\mathbb{R}\ra\mathbb{R}$.
The space of all $U_{\gamma}$ is denoted by $\mathbf{T}_{\gamma}\Gamma_{\gamma}$.
If necessary, $U_{\gamma}$ can be extended to a vertical vector field $\tilde{U}$ in an open neighbourhood of the image of $\gamma$ written as 
\be
\tilde{U}\,=\,U_{\varphi}\frac{\delta}{\delta \varphi} + U_{p}\frac{\delta}{\delta p} + U_{\beta }^{j} \frac{\delta}{\delta \beta^{j}},
\ee
where $U_{\varphi}\in\stav_\Sigma$, $U_{p}\in\mathscr{W}_\Sigma$, $U_{\beta}^{j}\in\mathscr{B}_\Sigma$. 
The symbols $\frac{\delta}{\delta \varphi}$ have a double interpretation.
On one side, they represent the canonical basis of tangent vectors on a linear space associated to the natural chart provided by the linear space itself.
On the other, they  represent the standard functional derivatives\cite{AbraMars-Foundations_of_Mechanics}, i.e., they act on a function $F$ as the functional derivative $\frac{\delta F}{\delta \varphi (x)}$, and similarly for $\frac{\delta}{\delta p (x)}$ and $\frac{\delta}{\delta \beta^{j}(x)}$.  
In a similar way, the symbols $\delta \varphi$ and $\delta p$ can be understood as a basis of covectors on the manifold $\mathcal{F}_\Sigma$, and $\delta F = \frac{\delta F}{\delta \sigma} \delta \sigma$ is understood as the differential of the function $F$ given by
$$
\langle \delta F, U_\sigma \rangle = \int_\Sigma \frac{\delta F}{\delta \sigma (x)} \,  U_\sigma (x) \,  \mathrm{vol}_\Sigma (x) \, ,
$$
Now,   we define the Hamiltonian function
\begin{equation}
\mathcal{H}(\varphi,p,\beta;s) = \int_{\Sigma} \dfrac{1}{2}\left( p^2 + \delta^{jk} \frac{\partial \varphi}{\partial x^{j}}\frac{\partial \varphi}{\partial x^k} - \mathfrak{m}^2 \varphi^2 \right)\mathrm{vol}_{\Sigma}\,=:\,\int_{\Sigma}\,H\,\mathrm{vol}_{\Sigma}\,,
\end{equation}
and we consider the one-form $ \mathcal{F}_{\Sigma}\times\mathbb{R}$ given by
\be\label{eqn: infinite theta}
\Theta_{\mathcal{H}}\,=\,\int_{\Sigma}\,\left(p\,\delta\varphi\right)\,\mathrm{vol}_{\Sigma} - \mathcal{H}\dd t\,.
\ee
This form plays a role analogous to that of the contact one-form on the extended phase space $\mathbf{T}^{*}\mathcal{Q}\times\mathbb{R}$ in the case of non-relativistic Hamiltonian dynamics, and to that of the contact one-form on the mass-shell for the relativistic particle considered in the companion letter \cite{C-DC-I-M-S-2020-01}.

Then,   we write the action functional $S$ on $\Gamma(\mathcal{F}_{\Sigma})$ given by
\be
S[\gamma]\,=\,\int_{\mathbb{R}}\,\gamma^{*}\Theta_{\mathcal{H}},
\ee
and we compute its variation along $U_{\gamma}$ as
\be
\mathrm{d}S_{\gamma}(U_{\gamma})\,=\,\int_{\mathbb{R}}\,\gamma^{*}\left(i_{\tilde{U}}\dd\Theta_{\mathcal{H}} + \dd(i_{\tilde{U}}\Theta_{\mathcal{H}})\right)\,.
\ee
Exploiting Stokes' theorem and the fact that $\partial\mathbb{R}=\emptyset$, we obtain
\be
\mathrm{d}S_{\gamma}(U_{\gamma})\,=\,\int_{\mathbb{R}}\,\gamma^{*}\left(i_{\tilde{U}}\dd\Theta_{\mathcal{H}}  \right).
\ee
Following the Schwinger-Weiss action principle, the dynamical trajectories are given by all those $\gamma$ such that 
\be
\mathbb{EL}_{\gamma}(U_{\gamma})\,:=\,\int_{\mathbb{R}}\,\gamma^{*}\left(i_{\tilde{U}}\dd\Theta_{\mathcal{H}}  \right)\,=\,0 \quad\forall\,U_{\gamma}\,\,\mathbf{T}_{\gamma}\Gamma_{\gamma}\,.
\ee
A direct computation shows that all such $\gamma$ must satisfy the constraint relations
\begin{equation}
\delta_{jk}\beta^k = \frac{\partial \varphi}{\partial x^j} , 
\label{constraints klein-gordon}
\end{equation}
and the ``evolution equations'' 
\be 
\frac{\dd \varphi}{\dd s} \,=   p\, ,\quad\,\frac{\dd p}{\dd s}  = - \Delta_{\Sigma}  \varphi  - \mathfrak{m}^2 \varphi .
\label{Cauchy-form klein gordon equation}
\ee 
The constraint conditions in Eq. \eqref{constraints klein-gordon} being linear determine a submanifold $\mathcal{C}\subset \mathcal{F}_{\Sigma}$ which is a linear subspace.
We define the trivial bundle $\tau_{\mathcal{C}}\colon\mathcal{C}\times\mathbb{R}\ra\mathbb{R}$, where $  \tau_{\mathcal{C}}$ is the projection on the second factor, and we denote by $\Gamma_{\mathcal{C}}$ the space of sections of this bundle.
Moreover, we write, with an evident abuse of notation, $\Theta_{\mathcal{H}}$ for the pullback to $\mathcal{C}\times\mathbb{R}$ of the one-form in equation \eqref{eqn: infinite theta}.
Then, it is clear that   Eq.\eqref{Cauchy-form klein gordon equation} may be read as defining the integral curves of the (densely defined) vector field
\be
X_{H} = \frac{\partial}{\partial t}  + \frac{\delta H}{\delta p}\,\frac{\delta}{\delta \varphi} - \frac{\delta H}{\delta \varphi}\,\frac{\delta}{\delta p}\,
\ee
which is in the kernel of the two-form 
\begin{equation}
\dd \Theta_{\mathcal{H}}\, =\, \int_{\Sigma}\left( \delta p \wedge \delta\varphi\right)\mathrm{vol}_{\Sigma} - \dd\mathcal{H} \wedge  \dd t \,. 
\end{equation}
This two-form plays the role of the contact two-form on $\mathbf{T}^{*}\mathcal{Q}\times\mathbb{R}$ in the case of non-relativistic Hamiltonian mechanics, and of the contact two-form on the mass-shell in the case of the relativistic particle considered in \cite{C-DC-I-M-S-2020-01}.
Eventually, we obtained the de Donder-Weyl equations of the free Klein-Gordon theory  as a Hamiltonian system on an infinite-dimensional manifold, in such a way that a section $\gamma$ satisfying Eq.\eqref{Cauchy-form klein gordon equation} provides a representation in terms of Cauchy data on $\Sigma$ of the points in $\mathcal{EL}_{\mathcal{M}}$. 
We denote by $\mathcal{EL}_{\mathcal{C}}$ the space of all $\gamma\in\Gamma_{\mathcal{C}}$ satisfying Eq.\eqref{Cauchy-form klein gordon equation}.

Just as it happens for non-relativistic Hamiltonian mechanics and for the relativistic particle\cite{C-DC-I-M-S-2020-01},  any vector field of the form $f\,X_{H}$, with $f$ a smooth, non-vanishing function on $\mathcal{C}\times\mathbb{R}$ is again in the kernel of $\dd \theta_{\mathcal{H}}$, and the support of its integral curves coincide with the support of the integral curves of $X_{H}$.
Accordingly, we may interpret the integral curves of $f\,X_{H}$ as reparametrizations of the dynamical trajectories.
In particular, the vector field $\Gamma_H$ satisfying $i_{\Gamma_H}\theta_\mathcal{H} = 1$ is called Reeb vector field.  
Under suitable regularity properties for $X_H$, the family of its integral curves   defines a regular foliation of the manifold $\mathcal{C}\times \mathbb{R}$, and   every point in the quotient manifold, say $\mathrm{Q}$, associated with the foliation can be identified with one and only one element in $\mathcal{EL}_{\mathcal{C}}$.

Accordingly, we may look at the space of smooth functions on $\mathrm{Q}$ as the subalgebra $C^{\infty}_H(\mathcal{C}\times \mathbb{R})\subset C^{\infty} (\mathcal{C}\times \mathbb{R})$ of smooth functions such that $\mathrm{L}_{X_H}f=0$.
Then, we may look for a bivector $\Lambda$ on $\mathcal{C}\times \mathbb{R}$ satisfying the relations
\be
\left[\Lambda,\Lambda\right]_{S}\,=\,2\,\Gamma_H\,\wedge\,\Lambda\,,\quad\,\mathrm{L}_{\Gamma_H}\Lambda\,=\,0,
\ee
where $[,]_{S}$ denotes the Schouten-Nijenhuis bracket, in order to define a Jacobi bracket\cite{AsoCiagliaDCosmoIbortMarmo2017-Covariant_Jacobi_brackets,C-C-M-2018} $[,]_{J}$ on $ C^{\infty} (\mathcal{C}\times \mathbb{R})$ by means of 
\be
[f,g]_{J}\,:=\,\Lambda(\mathrm{d}f,\mathrm{d}g) + f\,\mathrm{L}_{\Gamma_H}g - g\,\mathrm{L}_{\Gamma_H}f\,.
\ee
Recalling that $\Gamma_{H}$ annihilates the elements in $C^{\infty}_H(\mathcal{C}\times \mathbb{R})$, it is immediate to check that $C^{\infty}_H(\mathcal{C}\times \mathbb{R})$ becomes a Poisson subalgebra of $C^{\infty}(\mathcal{C}\times \mathbb{R})$ with respect to the Jacobi bracket defined above.
Upon  identifying   $\mathcal{EL}_{\mathcal{M}}$ with $\mathcal{EL}_{\mathcal{C}}$, and then $\mathcal{EL}_{\mathcal{C}}$ with  $\mathrm{Q}$, the Poisson bracket on $C^{\infty}_H(\mathcal{C}\times \mathbb{R})$ is precisely the Poisson bracket associated with the two-form  $\Omega$ in Eq.\eqref{eqn: omega on solutions of de Donder Weyl for KG}.
To explicitely write the Jacobi and Poisson bracket, we must be able to find an analogue of the generalized Darboux coordinates used in the particle case in the companion letter \cite{C-DC-I-M-S-2020-01}.
At this purpose, we first move from $\mathcal{C}$ to $\mathcal{C}'$ introducing the Fourier transforms
\be
\varphi(x)\,=\,\int_{\overline{\Sigma}}\,\hat{\varphi}(k)\,\mathrm{e}^{i k\cdot x}\,\,\mathrm{vol}_{\overline{\Sigma}},\quad\,p(x)\,=\,\int\,\hat{p}(k)\,\mathrm{e}^{i k\cdot x}\,\mathrm{vol}_{\overline{\Sigma}}.
\ee
Note that the fact that $\varphi$ and $p$ are real-valued  imposes the constraints $\overline{\hat{\varphi}}(k)=\hat{\varphi}(-k)$ and $\overline{\hat{p}}(k)=\hat{p}(-k)$.
The equations of motion become
\be
\frac{\dd \hat{\varphi}}{\dd s}\,=\,-\hat{p},\quad\,\frac{\dd\hat{p}}{\dd s}\,=\, \omega_{k}\,\hat{\varphi}\,,
\ee
with $\omega_{k}=(k^{2}+\mathfrak{m}^{2})$, and we may read them as a superposition in $k$ of harmonic oscillators.
Now, we consider the  ``change of coordinates'' in $\mathcal{C}'\times\mathbb{R}$ given by
\be
\begin{split}
W\,=\,\frac{1}{2}\int_{\overline{\Sigma}}\,\biggl( \left(\hat{p}\,\overline{\hat{p}} - \omega_{k}^{2}\,\hat{\varphi}\,\overline{\hat{\varphi}}\right) & \frac{\cos(\omega_{k}s)\,\sin(\omega_{k}s)}{\omega_{k}} + 2\left(\hat{p}\overline{\hat{\varphi}} + \overline{\hat{p}}\,\hat{\varphi}\right)\sin^{2}(\omega_{k}s) \biggr) \,\mathrm{vol}_{\overline{\Sigma}} \\
\hat{\Phi}&\,=\,\hat{\varphi}\,\cos(\omega_{k}s) - \frac{\hat{p}}{\omega_{k}}\,\sin(\omega_{k}s) \\
\hat{P}&\,=\,\hat{p}\cos(\omega_{k}s) + \omega_{k}\hat{\varphi}\,\sin(\omega_{k}s)\,,
\end{split}
\ee
where, again, it is $\overline{\hat{\Phi}}(k)=\hat{\Phi}(-k)$ and $\overline{\hat{P}}(k)=\hat{P}(-k)$.
In this coordinate system, it is possible to see that the one-form $\Theta_{\mathcal{H}}$ in equation \eqref{eqn: infinite theta} becomes
\be
\Theta_{\mathcal{H}}\,=\,\int_{\overline{\Sigma}} \,\frac{1}{2}\,\left(\overline{\hat{P}}\,\delta\hat{\Phi} + \hat{P}\,\delta\overline{\hat{\Phi}}\right) \,\mathrm{vol}_{\overline{\Sigma}} + \dd W\,.
\ee 
The   vector field  $\Gamma_{H}$ becomes $\Gamma_{H}\,=\,\frac{\partial}{\partial W}$, and, in analogy with what we did in the particle case in the companion letter \cite{C-DC-I-M-S-2020-01}, the bivector field $\Lambda$ in the definition of the Jacobi bracket reads
\be
\Lambda\,=\,\int_{\overline{\Sigma}}\,\left(\frac{\delta}{\delta \hat{\Phi} } - \hat{P} \,\frac{\partial}{\partial W}\right)\,\wedge\,\frac{\delta}{\delta \overline{\hat{P}} }\,\mathrm{vol}_{\overline{\Sigma}}  \,.
\ee
Clearly, elements in  $C^{\infty}_H(\mathcal{C}'\times \mathbb{R})$ are just those functionals which do not depend on $W$, and thus the Jacobi bracket among them   becomes the Poisson bracket given by
\be
\Lambda(\dd F,\dd G)\,=\,\int_{\overline{\Sigma}}\, \frac{\delta F}{\delta \hat{\Phi} } \,\wedge\,\frac{\delta G}{\delta \overline{\hat{P}} }\,\mathrm{vol}_{\overline{\Sigma}}  \,.
\ee

\section{Multisymplectic formulation of free Schr\"{o}dinger equation}

In this section, we will consider the Schr\"{o}dinger equation for a free particle\footnote{The case of unitary evolutions of a finite-level quantum system\cite{C-DC-I-L-M-2017,C-DC-L-M-2017} may be consistently dealt with within the formalism described in the companion letter \cite{C-DC-I-M-S-2020-01}.} in $\mathbb{R}^{3}$.
The analysis of the Schr\"{o}dinger equation with a potential may be given following the lines presented here, bearing in mind, however, that the presence of the potential may affect the choice of the appropriate Sobolev and Hilbert spaces for the fields considered.

For the Schr\"{o}dinger equation, our spacetime manifold will be $\mathcal{M}\cong\mathbb{R}^{4}$.
In this case, we are in a non-relativistic context, where the notion of {\itshape absolute simultaneity} is available\cite{deritis_marmo_preziosi-a_new_look_at_relativity_transformations,MarmoPreziosi}, and thus the spacetime may be splitted in the product of space and time.
The notion of absolute simultaneity is encoded in the existence of an exact one-form $\vartheta$ on $\mathcal{M}$ the kernel of which determines an integrable foliation whose leaves are diffeomorphic with $\mathbb{R}^{3}$.
The foliation associated with $\vartheta$ defines absolute simultaneity, and the leaves define the simultaneity surfaces.
Since $\vartheta$ is exact, there will be a global time function $t\colon\mathcal{M}\ra\mathbb{R}$ such that $\vartheta\,=\,\dd t$, and the simultaneity surfaces are identified with the level sets of the time function.
In the following,  we will always choose Cartesian coordinates $(x^{1},x^{2},x^{3},t)$ reflecting the fact that the $t$-coordinate has a clear and definite physical interpretation being the global time function defining absolute simultaneity.
A simultaneity surface will be denoted by $\Sigma_{\bar{t}}$, where $\bar{t}$ is the value of the time function characterizing the simultaneity surface.

Differently from Klein-Gordon theory, the wave function $\psi$ of Schr\"{o}dinger equation is a complex-valued function on $\mathcal{M}$, which in addition does not transform as a function but as a section of a $U(1)$-bundle \cite{ECM}. 
However, we will describe $\psi$ in a real-valued context by considering a pair of fields $(\phi^R, \phi^I)$ associated with any wave function, where $\phi^{R}$ denote the real part of the wave function, and $\phi^{I}$ the imaginary part.
Any field will have its own momenta, say $(P^{0}_R, P^{j}_R)$ and $(P^{0}_I, P^{j}_I)$, and the covariant phase space is $\pi\colon\mathcal{P}=\mathbb{R}^{10}\times \mathcal{M}\rightarrow \mathcal{M}$, where $\pi$ is the projection on the second factor, which is a vector bundle over $\mathcal{M}$ as in the case of Klein-Gordon theory.
Coordinate functions on $\mathcal{P}$ are $(u^a,\rho^{j}_a;\rho^{0}_{a}, x^{j},t)$, with $a=R,I$ and $j= 1,2,3$. 

A reference frame on $\mathcal{M}$ is defined by the choice of a nowhere-vanishing vector field $\Gamma$ on $\mathcal{M}$ such tha $\vartheta(\Gamma)=1$ \cite{AuchmannKurz,deritis_marmo_preziosi-a_new_look_at_relativity_transformations,Fecko,MarmoPreziosi, CiagliaDCMarmoSchiavone}.
As we are dealing with Schr\"{o}dinger equation, it seems natural to limit ourselves to inertial Galilean frames.
These are those reference frames characterized by $\Gamma=\frac{\partial}{\partial t} + v^{j}\frac{\partial}{\partial x^{j}}$.
Given $\Gamma$, we can complete it to a frame of vector fields on $\mathcal{M}$ by introducing the vector fields $\frac{\partial}{\partial x^{j}}$ with $j=1,2,3$.
These ``auxiliary'' vector fields are taken to be in the kernel of $\vartheta=\dd t$ so that they are tangent to the simultaneity surfaces determined by $\vartheta=\dd t$.
Then, we determine the associated dual frame given by $\{\vartheta=\dd t,\,\dd x^{j} - v^{j}\dd t\}$ with $j=1,2,3$ and we build the frame-dependent covariant tensor $G_{\Gamma}$ given by
\be
G_{\Gamma}\,=\,\delta_{jk}\,\left(\dd x^{j} - v^{j}\dd t\right)\otimes \left(\dd x^{k} - v^{k}\dd t \right).
\ee
It is straightforward to notice that it determines an Euclidean metric tensor on every simultaneity leaf associated with $\vartheta=\dd t$.
Now, by means of $G_{\Gamma}$ we can define the frame-dependent Hamiltonian function given by
\be
H_{\Gamma}\,=\,-\frac{\delta_{ab}}{2}\left(\delta_{jk}v^{j}v^{k}\rho^{0}_{a}\rho^{0}_{b} - \delta_{jk}v^{k}(\rho^{j}_{a}\rho^{0}_{b} + \rho^{j}_{b}\rho^{0}_{a}) + \delta_{jk}\rho^{j}_{a}\rho^{k}_{b}\right).
\ee
We stress that, in the non-relativistic case, the Hamiltonian function used for the description of a free quantum particle depends on the choice of a reference frame on $\mathcal{M}$.
This is due to the fact that the covariant tensor $G_{\Gamma}$ is only defined once we   determine the frame of vector fields $\{\Gamma,\frac{\partial}{\partial x^{j}}\}$, with $j=1,2,3$, and its dual frame $\{\vartheta, \,\dd x^{j} - v^{j}\dd t\}$ with $j=1,2,3$.
On the other hand, in the relativistic case, we already have the spacetime Lorentzian metric $\eta$ and we do not need the choice of a reference frame to define the Hamiltonian function (see equation \eqref{eqn: relativistic Hamiltonian}).

In the following, we are going to perform our analysis in the inertial reference frame identified by $\vartheta=\dd t$ and $\Gamma=\frac{\partial}{\partial t}$,  so that $G_{\Gamma}\,=\,\delta_{jk}\dd x^{j}\otimes\dd x^{k}$, and the Hamiltonian function reads
\be
H = -\delta_{ab}\delta_{jk} \frac{\rho^{j}_{a}\rho^{k}_{b} }{2},
\ee
where we have set $H_{\Gamma}\equiv H$.
However, everything that will be said below could be adapted to a different choice for $\Gamma$ thus obtaining a description in a different inertial Galilean reference frame. Now, we can define the  4-form $\tilde{\theta}_H$ on $\mathcal{P}$ given by
\begin{equation}
\tilde{\theta}_H = \rho^{0}_a \dd u^a\wedge i_{\Gamma}\mathrm{vol}_\mathcal{M}  + \rho^{j}_a \dd u^a\wedge i_{\frac{\partial}{\partial x^{j}}}\mathrm{vol}_\mathcal{M}    - H\mathrm{vol}_\mathcal{M}  \, .
\label{4-form schrodinger}
\end{equation}  
Note that the vector field $\Gamma$ defines  the Galielan reference frame, while the vector fields $\frac{\partial}{\partial x^{j}}$ transform as the component of a vector with respect to the action of the  Galilei group.
Unlike the relativistic case dealt with in section \ref{sec: KG}, the 4-form  $\tilde{\theta}_H$  on $\mathcal{P}$ depends on the choice of a reference frame on $\mathcal{M}$ because the Hamiltonian function does so.
To be able to recover Schr\"{o}dinger equation, we must impose two constraints and select a sub-bundle of $\mathcal{P}$.
Specifically, we consider the vector sub-bundle  $\pi\colon\mathcal{Q}\ra\mathcal{M}$ singled out by the constraints 
\be 
\rho^0_R - u^I = 0 ,\quad \rho^0_I + u^R = 0\,.
\ee 
The origin of these constraints is related to the circumstance that  wave-functions under Galilei transformations behave like sections of a $U(1)$ bundle  \cite{ECM}.
Moreover, they appear naturally when the Schr\"{o}dinger equation is written by means of a suitable reduction procedure of a 5-dimensional equation.
This aspect will be considered elsewhere. Let $\chi$ be a section of  $\pi\colon\mathcal{Q}\ra\mathcal{M}$ given by
\begin{equation}
\chi(x^{\mu}) = (x^{j},t, \phi^{R}(x^{j},t), P^{k}_{R}(x^{j},t),\phi^{I}(x^{j},t), P^{k}_{I}(x^{j},t))\,.
\end{equation}
As before, we assume some regolarity conditions on the sections we actually consider.
Specifically, we take $\left( \phi^a \right) \in \mathcal{V}=\mathcal{H}^1(\mathcal{M}, \mathrm{vol}_M)$ and $\left( P^{j}_a \right) \in \mathcal{W}= \mathcal{L}^2(\mathcal{M}, \mathrm{vol}_M)$.
In this way, the space of fields of the theory is  the Hilbert space $\mathscr{F}_{\mathcal{Q}}\,=\,\mathcal{V}\oplus\mathcal{W}$.
A variation for $\chi\in\mathscr{F}_{\mathcal{Q}}$ is a tangent vector at  $\chi$, which may be identified witha vector field $U_{\chi}$ along along $\chi$ on $\mathcal{Q}$, which is vertical with respect to the fibration $\pi\colon\mathcal{Q}\ra\mathcal{M}$.
The tangent space $\mathbf{T}_{\chi}\mathscr{F}_{\mathcal{Q}}$ is given by all the $U_{\chi}$.
In the following, it will be useful to extend $U {\chi}$ to a vertical vector field $\widetilde{U}$ in a neighbourhood of the image of $\chi$ inside $\mathcal{P}$ given by
\begin{equation}
\tilde{U} = U^{\phi}_{R} \frac{\partial }{\partial u^{R}} +        U_{R}^{j}\frac{\partial }{\partial \rho^{j}_{R}} +  U^{\phi}_{I} \frac{\partial }{\partial u^{I}} +    U_{I}^{j}\frac{\partial }{\partial \rho^{j}_{I}}\,.
\end{equation}

As before, the dynamics may be described in terms of the Schwinger-Weiss action principle for sections $\chi\in\mathscr{F}_{\mathcal{Q}}$.
First of all, we consider  the pullback $\theta_H $  to $\mathcal{Q}$ of the form $\tilde{\theta}_H $ in Eq.\eqref{4-form schrodinger} given by
\begin{equation}
\theta_H = \left( u^I\dd u^R - u^R\dd u^I \right)\wedge i_{\Gamma}\mathrm{vol}_{\mathcal{M}} + \rho^j_a \dd u^a \wedge i_{\frac{\partial}{\partial x^{j}}}\mathrm{vol}_\mathcal{M}  - H\mathrm{vol}_\mathcal{M}  \, .
\end{equation}
Then, we    define the action functional $S$ on $\mathscr{F}_{\mathcal{Q}}$ as  
\begin{equation}
S[\chi]= \int_{\mathcal{M}}\chi^*\left( \theta_H \right)\,, 
\end{equation}
with $\chi\in\mathscr{F}_{\mathcal{Q}}$.
At this point, we may proceed as in the previous section  and compute the  variation  $\mathrm{d}S[\chi](U_{\chi})$ of $S$ to be
\be
\begin{split}
\mathrm{d}S[\chi](U_{\chi})&=  \int_{\mathcal{M} }\chi^*\left( \mathrm{L}_{\tilde{U}}\theta_H \right)  = \int_{\mathcal{M} }\chi^*\left( i_{\tilde{U}}\dd \theta_H \right) + \int_{\mathcal{M} } \dd \chi^*\left( i_{\tilde{U}}\theta_H \right),
\end{split}
\ee
where $\tilde{U}$ is any extension of $U_{\chi}$.
Again, we use Stokes' theorem and see that the boundary term vanishes because $\partial\mathcal{M}=\emptyset$.
The action principle  states that the dynamical trajectories satisfy the Euler-Lagrange equations for the action functional
\begin{equation}\label{eqn: Schwinger-Weiss action principle 2}
\mathbb{EL}_{\chi}(U_{\chi}) := \int_{\mathcal{M} }\chi^*\left( i_{\tilde{U}}\dd \theta_H \right) = 0 \,,\quad \forall U_{\chi}\in\mathbf{T}_{\chi}\mathscr{F}_{\mathcal{P}}\,,
\end{equation}
which are nothing but the de Donder-Weyl equations
\begin{equation}
\begin{split}
\frac{\partial \phi^I}{\partial t} = - \frac{1}{2}\frac{\partial P^j_R}{\partial x^j} \,, \qquad  \frac{\partial \phi^I}{\partial x^j} = - \delta_{jk}P^k_I  \\
\frac{\partial \phi^R}{\partial t} = \frac{1}{2}\frac{\partial P^j_R}{\partial x^j} \,, \qquad \frac{\partial \phi^R}{\partial x^j} = - \delta_{jk}P^k_R\,.
\end{split}
\label{deDonder-Weyl-equations Schrodinger}
\end{equation}
It is a matter of straightforward computation to see that the free Schr\"{o}dinger equation for $\psi$ follows from equation \eqref{deDonder-Weyl-equations Schrodinger} upon writing $\psi= \phi^R + i\phi^I$.
The space of solutions is denoted by $\mathcal{EL}_{\mathcal{M}}$, and it is equipped with the two-form 
\begin{equation}\label{eqn: omega on solutions of de Donder Weyl for Schroedinger}
\Omega^{\Sigma}_{\chi}(U_{\chi},V_{\chi}) = \int_{\Sigma}i_{\Sigma}^*\chi^*(i_{\tilde{V}}i_{\tilde{U}}\dd \theta_H)\,.
\end{equation}
Once again, following the steps outlined in the companion letter \cite{C-DC-I-M-S-2020-01}, it is possible to show that $\Omega^{\Sigma}$ is actually independent of the simultaneity surface 
and frame of reference, and we can simply write $\Omega$. 

In the remainder of the section, we want to formulate the dynamics in terms of an action principle for sections of an infinite-dimensional bundle which replaces $\mathcal{Q}$.
In this way, we will be able to read the bracket associated with $\Omega$ in terms of a Poisson subalgebra of the algebra of smooth functions on a suitable   manifold endowed with a Jacobi bracket.
We proceed in analogy with what is done in the previous section.
Therefore, we fix a time-slice $\Sigma$ and we consider the pullback bundle $i_{\Sigma}^{*}\mathcal{Q}$, where $i_{\Sigma}$ is the immersion map of $\Sigma$ inside $\mathcal{M}$.
A section $\sigma$ of this bundle is given by
$$
\sigma(x^{j})=\left(x^{j},t^{0}, \varphi(x^{j})^{a},\,   \beta^k_{a}(x^{j})  \right),
$$
where $j=1,2,3$ and $a=R,I$, and where
\be
\varphi^{a}(x^{j})=(\phi^{a}(x^{j},t^{0})) |_{\Sigma},\quad \beta^k_{a}(x^{j})=(P^{k}_{a}(x^{j},t^{0})) |_{\Sigma} \,,
\ee 
for some section $\chi(x^{j},t)\,=(x^{j},t,\phi^{a}(x^{j},t), P^{k}(x^{j},t)$ in $\mathscr{F}_{\mathcal{Q}}$.
As before, we impose some regularity conditions on the admissible sections $\chi$.
Specifically,   we will consider $(\varphi^{a}) \in\stav=\mathcal{H}^1(\Sigma, \mathrm{vol}_{\Sigma})$,  and $(\beta^{j}_{a}) \in\mathscr{B}=\mathcal{L}^2(\Sigma, \mathrm{vol}_{\Sigma})$, so that we have the Hilbert space of fields   given by $\mathcal{F}_{\Sigma}\,:=\,\stav\oplus \mathscr{B}$.
Then, we consider the Hilbert bundle $\tau\colon\mathcal{F}_{\Sigma}\times\mathbb{R}\ra\mathbb{R}$, where $\tau$ is the projection on the second factor, which plays the role of the extended phase space in \cite{C-DC-I-M-S-2020-01}, and we denote by $\Gamma$ the space of sections of this bundle.
Following what is done in the previous section, we can find a one-to-one correspondence between $\Gamma$ and $\mathscr{F}_{\mathcal{Q}}$. Now, we define the Hamiltonian function
\begin{equation}
\mathcal{H}(\varphi^{a},\beta^{j}_{a};s) = \int_{\Sigma} -\frac{1}{2}\delta^{jk}\left( \frac{\partial \varphi^R}{\partial x^j}\frac{\partial \varphi^R}{\partial x^k} + \frac{\partial \varphi^I}{\partial x^j}\frac{\partial \varphi^I}{\partial x^k} \right) \mathrm{vol}_{\Sigma}=:\int_{\Sigma}\,H\,\mathrm{vol}_{\Sigma}\,,
\end{equation}
the one-form
\begin{equation}\label{eqn: infinite theta2}
\Theta_{\mathcal{H}}=2\int_{\Sigma} \varphi^I \delta\varphi^R \mathrm{vol}_{\Sigma} -   \mathcal{H}\wedge \dd t\,,
\end{equation}
and the action functional
\be
S[\gamma]=\int_{\mathbb{R}}\,\gamma^{*}\Theta_{\mathcal{H}}.
\ee
Developing the variation of $S$ as done for the Klein-Gordon case, we obtain the constraints
\be\label{constraints schroedinger}
\frac{\partial \varphi^R }{\partial x^j} = - \delta_{jk}\beta^k_R\,, \quad \frac{\partial \varphi^I }{\partial x^j} = - \delta_{jk}\beta^k_I 
\ee
and the ``evolution equations''
\be \label{Cauchy-form schroedinger equation}
\frac{\dd \varphi^R}{\dd s} = - \frac{1}{2} \Delta_{\Sigma}\varphi^I , \quad \frac{\dd \varphi^I}{\dd s} = \frac{1}{2} \Delta_{\Sigma}\varphi^R \,.
\ee
The constraint conditions in Eq.\eqref{constraints schroedinger} determine a submanifold $\mathcal{C}\subset \mathcal{F}_{\Sigma}$ which is a linear subspace.
We define the trivial bundle $\tau_{\mathcal{C}}\colon\mathcal{C}\times\mathbb{R}\ra\mathbb{R}$, where $  \tau_{\mathcal{C}}$ is the projection on the second factor, and we denote by $\Gamma_{\mathcal{C}}$ the space of sections of this bundle.
Moreover, we write, with an evident abuse of notation, $\Theta_{\mathcal{H}}$ for the pullback to $\mathcal{C}\times\mathbb{R}$ of the one-form in equation \eqref{eqn: infinite theta2}.

Then, on $\mathcal{C}\times\mathbb{R}$, it is clear that   Eq.\eqref{Cauchy-form schroedinger equation} may be read as defining the integral curves of the (densely defined) vector field
\be
X_{H} = \frac{\partial}{\partial t}  + \frac{\delta H}{\delta \varphi^{I}}\,\frac{\delta}{\delta \varphi^{R}} - \frac{\delta H}{\delta \varphi^{R}}\,\frac{\delta}{\delta \varphi^{I}}\,
\ee
which is in the kernel of the two-form 
\begin{equation}
\dd \Theta_{\mathcal{H}}\, =\, 2\int_{\Sigma}\left( \delta \varphi^{I} \wedge \delta\varphi^{R}\right)\mathrm{vol}_{\Sigma} - \dd\mathcal{H} \wedge  \dd t \,.
\end{equation}
This two-form plays the role of the contact two-form on $\mathbf{T}^{*}\mathcal{Q}\times\mathbb{R}$ in the case of non-realtivistic Hamiltonian mechanics, and of the contact two-form on the mass-shell in the case of the relativistic particle considered in \cite{C-DC-I-M-S-2020-01}.
Eventually, we obtained the de Donder-Weyl equations for the free quantum particle in $\mathbb{R}^{3}$  as a Hamiltonian system on an infinite-dimensional manifold, in such a way that a section $\gamma$ satisfying Eq.\eqref{Cauchy-form schroedinger equation} provides a representation in terms of Cauchy data on $\Sigma$ of the points in $\mathcal{EL}_{\mathcal{M}}$. 
We denote by $\mathcal{EL}_{\mathcal{C}}$ the space of all $\gamma\in\Gamma_{\mathcal{C}}$ satisfying Eq.\eqref{Cauchy-form schroedinger equation}.

Again, all vector fields $fX_{H}$ with $f$ a non-vanishing function are in the kernel of $\dd\Theta_{\mathcal{H}}$, and the integral curves of $f\,X_{H}$ can be interpreted as reparametrizations of the dynamical trajectories.
Moreover, just as we said for the free Klein-Gordon theory,  under suitable regularity properties for $X_H$, the family of its integral curves   defines a regular foliation of the manifold $\mathcal{C}\times \mathbb{R}$, and   every point in the quotient manifold, say $\mathrm{Q}$, associated with the foliation can be identified with one and only one element in $\mathcal{EL}_{\mathcal{C}}$.
This allows us to look at the space of smooth functions on $\mathrm{Q}\cong\mathcal{EL}_{\mathcal{C}}\cong\mathcal{EL}_{\mathcal{M}}$ as the subalgebra $C^{\infty}_H(\mathcal{C}\times \mathbb{R})\subset C^{\infty} (\mathcal{C}\times \mathbb{R})$ of smooth functions such that $\mathrm{L}_{X_H}f=0$.
Then, as we did for the Klein-Gordon equation, we may look for a bivector $\Lambda$ on $\mathcal{C}\times \mathbb{R}$ satisfying the relations
\be
\left[\Lambda,\Lambda\right]_{S}\,=\,2\,\Gamma_{H}\,\wedge\,\Lambda\,,\quad\,\mathrm{L}_{\Gamma_{H}}\Lambda\,=\,0,
\ee
where $[\cdot , \cdot ]_{S}$ denotes the Schouten-Nijenhuis bracket and $\Gamma_H$ is the Reeb vector field satisfying $i_{\Gamma_H}\Theta_{\mathcal{H}}=1$. The associated Jacobi bracket\cite{AsoCiagliaDCosmoIbortMarmo2017-Covariant_Jacobi_brackets,C-C-M-2018} $[\cdot , \cdot ]_{J}$ on $ C^{\infty} (\mathcal{C}\times \mathbb{R})$ is then defined as follows 
\be
[f,g]_{J}\,:=\,\Lambda(\mathrm{d}f,\mathrm{d}g) + f\,\mathrm{L}_{\Gamma_{H}}g - g\,\mathrm{L}_{\Gamma_{H}}f\,.
\ee
Recalling that $\Gamma_{H}$ annihilates the elements in $C^{\infty}_H(\mathcal{C}\times \mathbb{R})$, it is immediate to check that $C^{\infty}_H(\mathcal{C}\times \mathbb{R})$ becomes a Poisson subalgebra of $C^{\infty}(\mathcal{C}\times \mathbb{R})$ with respect to the Jacobi bracket defined above.
Upon  identifying   $\mathcal{EL}_{\mathcal{M}}$ with $\mathcal{EL}_{\mathcal{C}}$, and then $\mathcal{EL}_{\mathcal{C}}$ with  $\mathrm{Q}$, the Poisson bracket on $C^{\infty}_H(\mathcal{C}\times \mathbb{R})$ is precisely the Poisson bracket associated with the two-form  $\Omega$ in Eq. \eqref{eqn: omega on solutions of de Donder Weyl for Schroedinger}.
To explicitely write the Jacobi and Poisson bracket, we can look for an analogue of the generalized Darboux coordinates used in the particle case in the companion letter \cite{C-DC-I-M-S-2020-01}.
At this purpose, we first move from $\mathcal{C}$ to $\mathcal{C}'$ introducing the Fourier transforms
\be
\varphi_{R}(x)\,=\,\int_{\overline{\Sigma}}\,\hat{\varphi}_{R}(k)\,\mathrm{e}^{i k\cdot x}\,\,\mathrm{vol}_{\overline{\Sigma}}, \quad \varphi_{I}(x)\,=\,\int_{\overline{\Sigma}}\,\hat{\varphi}_{I}(k)\,\mathrm{e}^{i k\cdot x}\,\,\mathrm{vol}_{\overline{\Sigma}}.
\ee
Note that the fact that $\varphi_{R}$ and $\varphi_{I}$ are real-valued  imposes the constraints $\overline{\hat{\varphi}}_{R}(k)=\hat{\varphi}_{R}(-k)$ and $\overline{\hat{\varphi}}_{I}(k)=\hat{\varphi}_{I}(-k)$.
Now, we consider the  ``change of coordinates'' in $\mathcal{C}'\times\mathbb{R}$ given by
\be
\begin{split}
W\,=\, \frac{1}{2}\,\int_{\overline{\Sigma}}\  \left( |\hat{\Phi}_{R}|^{2} - |\hat{\Phi}_{I}|^{2}\right)   &\sin(k^{2}s) + 2 (\hat{\Phi}_{R}\overline{\hat{\Phi}}_{I} + \overline{\hat{\Phi}}_{R}\hat{\Phi}_{I})\sin\left(\frac{ k^{2}}{2} s\right)   \,\mathrm{vol}_{\overline{\Sigma}} \\
\hat{\Phi}_{R} &\,=\, \hat{\varphi_{R}}\,\cos\left(\frac{k^{2}}{2}s\right) -   \hat{\varphi_{I}}\,\sin\left(\frac{k^{2}}{2}s\right)\\
\hat{\Phi}_{I} &\,=\, \hat{\varphi_{I}}\,\cos\left(\frac{k^{2}}{2}s\right) + \hat{\varphi_{R}}\,\sin\left(\frac{k^{2}}{2}s\right)   ,.
\end{split}
\ee
where, again, it is $\overline{\hat{\Phi}}_{R}(k)=\hat{\Phi}_{R}(-k)$ and $\overline{\hat{\Phi}}_{I}(k)=\hat{\Phi}_{I}(-k)$.
In this coordinate system, it is possible to see that the one-form $\Theta_{\mathcal{H}}$ in equation \eqref{eqn: infinite theta2} becomes
\be
\Theta_{\mathcal{H}}\,=\,2\int_{\overline{\Sigma}} \,\overline{\hat{\Phi}}_{I} \,\delta\hat{\Phi}_{R}   \,\mathrm{vol}_{\overline{\Sigma}} +  \dd W\,.
\ee 
The Reeb vector field  $\Gamma_{H}$ becomes $\Gamma_{H}\,=\,\frac{\partial}{\partial W}$, and, in analogy with what we did in the particle case in the companion letter \cite{C-DC-I-M-S-2020-01}, the bivector field $\Lambda$ in the definition of the Jacobi bracket reads
\be
\Lambda\,=\,\frac{1}{2}\int_{\overline{\Sigma}}\,\left(\frac{\delta}{\delta \hat{\Phi}_{R} } - \hat{\Phi}_{I} \,\frac{\partial}{\partial W}\right)\,\wedge\,\frac{\delta}{\delta \overline{\hat{\Phi}}_{I} }\,\mathrm{vol}_{\overline{\Sigma}}  \,.
\ee
Clearly, elements in  $C^{\infty}_H(\mathcal{C}'\times \mathbb{R})$ are just those functions which do not depend on $W$, and thus the Jacobi bracket among them becomes the Poisson bracket given by
\be
\Lambda(\dd F,\dd G)\,=\,\frac{1}{2}\int_{\overline{\Sigma}}\, \frac{\delta F}{\delta \hat{\Phi}_{R}  } \,\wedge\,\frac{\delta G}{\delta \overline{\hat{\Phi}}_I }\,\mathrm{vol}_{\overline{\Sigma}}  \,.
\ee

\section{Conclusions}

In this letter, we continued the analysis of the description of the covariant bracket on the space of functionals on solutions to variational problems in the framework of contact geometry initiated in the companion letter \cite{C-DC-I-M-S-2020-01}.
We analysed in detail the case of free Klein-Gordon theory on Minkowski spacetime, and of the free Schr\"{o}dinger equation for a particle in $\mathbb{R}^{3}$.
Both systems were described by means of  two different but equivalent formulations.
On the one hand, we exploited the multisymlectic formalism to give a description in which the fields of the theory are sections of a suitable bundle over the spacetime manifold,  but for which it is not possible to appreciate the role of contact geometry.
Indeed, this formulation seems to point at the need to develop a generalization of contact geometry which is analogue to the multisymplectic generalization of symplectic geometry, and that will be pursued in a future work.

On the other hand, we presented a description in terms of a vector field on a suitable infinite-dimensional  manifold that leads to the identification  of the covariant bracket in terms of a Poisson subalgebra of the algebra of smooth functions on the infinite-dimensional manifold endowed with the Jacobi bracket.

\section*{Acknowledgments}

F.D.C. and A.I. would like to thank partial support provided by the MINECO research project MTM2017-84098-P and QUITEMAD++, S2018/TCS-A4342. A.I. and G.M. acknowledge financial support from the Spanish Ministry of Economy and Competitiveness, through the Severo Ochoa Programme for Centres of Excellence in RD(SEV-2015/0554). G.M. would like to thank the support provided by the Santander/UC3M Excellence Chair Programme 2019/2020, and he is also a member of the Gruppo Nazionale di Fisica Matematica (INDAM), Italy.

%

\end{document}